\documentclass[preprint,numbers,sort&compress]{elsarticle}
\usepackage[utf8]{inputenc}

\usepackage{mathtools}
\usepackage{hyperref}
\usepackage{natbib}
\usepackage{graphicx}
\title{On the development of an original mesoscopic model to predict the capacitive properties of carbon-carbon supercapacitors}

\author{Anouar Belhboub$^{1,2}$}
\ead{belhboub@chimie.ups-tlse.fr}

\author{El Hassane Lahrar$^{1,2}$}
\ead{lahrar@chimie.ups-tlse.fr}

\author{Patrice Simon$^{1,2}$}
\ead{simon@chimie.ups-tlse.fr}

\author{\\C\'eline Merlet$^{1,2,}$\corref{cor1}}
\ead{merlet@chimie.ups-tlse.fr}

\address{$^1$ \small CIRIMAT, Universit\'e de Toulouse, CNRS, France}

\address{$^2$ R\'eseau sur le Stockage \'Electrochimique de l'\'Energie (RS2E), F\'ed\'eration de Recherche CNRS 3459, HUB de l'\'Energie, Rue Baudelocque,  80039 Amiens, France}

\cortext[cor1]{Corresponding author: merlet@chimie.ups-tlse.fr}

\date{}

\begin{document}

\begin{abstract}
We report on the development of an original mesoscopic lattice model to predict structural, dynamical and capacitive properties of carbon-carbon supercapacitors. The model uses input from molecular simulations, such as free energy profiles to describe the ion adsorption, and experiments, such as energy barriers for transitions between lattice sites. The model developed is approximately 10,000 times faster than common molecular simulations. We apply this model to a set of carbon structures with well-defined pore sizes and investigate the solvation effect by doing simulations with neat ionic liquids as well as acetonitrile-based electrolytes. We show that our model is able to predict quantities of adsorbed ions and capacitances in a range compatible with experimental values. We show that there is a strong dependency of the calculated properties on the pore size and on the presence or absence of solvent. In particular, for neat ionic liquids, larger capacitances are obtained for smaller pores, while the opposite trend is observed for organic electrolytes.
\end{abstract}


\maketitle

\section{Introduction}

Carbon-carbon supercapacitors are electrochemical energy storage systems which store energy through ion adsorption at the interface between an electrolyte and porous carbon electrodes. Their high power density and long cycle life make them attractive for a number of applications in which they complement or sometimes even replace batteries. Supercapacitors are already used in applications such as regenerative energy braking~\cite{Yoshida17} and catenary-free trams~\cite{Zhang17c} where they are charged at every stop of the vehicle. Nevertheless, their relatively low energy density ($<$~20~Wh~kg$^{-1}$) compared to the one of batteries ($>$~150~Wh~kg$^{-1}$)~\cite{Simon2008,Guan2017} still limits their range of applications. 

In 2006, there was a breakthrough in the field of supercapacitors when it was demonstrated that electrolyte ions could enter pores of subnanometer sizes leading to a large capacitance (and thus energy density) increase~\cite{Chmiola06,Raymundo-Pinero06}. Since then, a number of experimental and theoretical projects have focused on understanding this capacitance increase and on designing new electrode materials with improved performances~\cite{Sheberla2017,Zhu2011,Zhang2009,Tao2013,Simon2013,Merlet2012,Liao2015}. While the capacitance increase is now well understood~\cite{Merlet2013a,Salanne2016}, reports of improvements in terms of capacitance are limited. Recent works report a capacitance of approximately 200~F~g$^{-1}$ in mesoporous~\cite{Liu2016} and pillared graphene~\cite{Banda19b} based supercapacitors.

One important issue in the field of supercapacitors is to assess the maximum capacitance that we could theoretically reach with an optimum system. Estimating a maximum capacitance in supercapacitors is a real challenge as it depends strongly on the ionic arrangements in the pores at a given potential which is the result of a large number of ion-ion and ion-electrode interactions. In supercapacitors, both the electrode and the electrolyte are disordered which makes them very difficult to characterise. A maximum value of 550~F~g$^{-1}$, often quoted, was proposed by Xia~\emph{et~al.}~\cite{Xia09}. This capacitance is the one which would be obtained for a single layer of counterions on both sides of a graphene layer, more precisely located 0.3~nm away from the carbon surface. The small ion-carbon distance considered and the fact that a full charge-separation between counter- and co-ions is assumed suggest that this value cannot be reached. However, there is to-date, no better estimation for the maximum capacitance reachable in electrochemical double layer capacitors. 

From a theoretical point of view, the most accurate approach to predict a capacitance is Molecular Dynamics simulations (MDs) because it allows one to describe the electrode-electrolyte interface microscopically~\cite{Merlet2012,Feng2011}. This accuracy comes with the price of a high computational cost which prevents the use of such a method to realise a screening of porous carbons for supercapacitor applications. Another issue with MDs is the small time and length scales which can be probe (a few nanoseconds, a few nanometers) usually far from the experimental values (a few micrometers at least, a few milliseconds or seconds). As such, the description of the carbon structure is usually not fully representative of the experimental reality, e.g. the pore size distributions of the model carbon electrodes are usually quite different from the experimental ones. Thus, in such systems, where phenomena on the atomistic scale have consequences on macroscopic length and timescales, it is important to develop new models to bridge the gap between molecular simulations and macroscopic values.

In this work, we report on the development of an original mesoscopic model to predict electrochemical performances of carbon-carbon supercapacitors at a much lower computational cost ($\sim$10$^4\times$ faster) than MD simulations. We adapt a lattice model which showed promising results for the simulation of ion diffusion and NMR spectra prediction for species adsorbed in neutral porous carbons~\cite{Merlet15,Forse15b}. Here, we implement new features to introduce the possibility of applying a potential to the modelled electrodes, and modify the way diffusion is treated in an attempt to get closer to reality. In the remainder of this article, we first describe the model before presenting the results obtained for two pure ionic liquids and two acetonitrile-based electrolytes in contact with a range of porous carbons having a simple pore size distribution (unimodal or bimodal). We study in particular the effects of solvation and pore size on the quantities of adsorbed ions and capacitive properties.

\section{Description of the lattice gas model}

In the lattice model, only one carbon electrode is simulated and its structure is represented as a cubic tridimensional set of inter-connected discrete sites, separated by a lattice spacing, $a$. Each lattice site is an accessible space represented by a slit pore in the porous matrix. The lattice pores are characterised by i) a pore size, i.e. the width of the pore, and ii) a pore surface, corresponding to the lateral walls of the pore. The pore sizes are assigned randomly across the matrix according to pore size distributions (PSD) obtained experimentally or from atomistic structures of different types of carbons (microporous, mesoporous,...). The pore surfaces are determined following a lognormal distribution with a mean value of -0.1 and a standard deviation of 0.25 (see reference~\cite{Merlet15} for a detailed explanation of this choice).

\begin{figure}[ht!]
\centering
\includegraphics[scale=0.2]{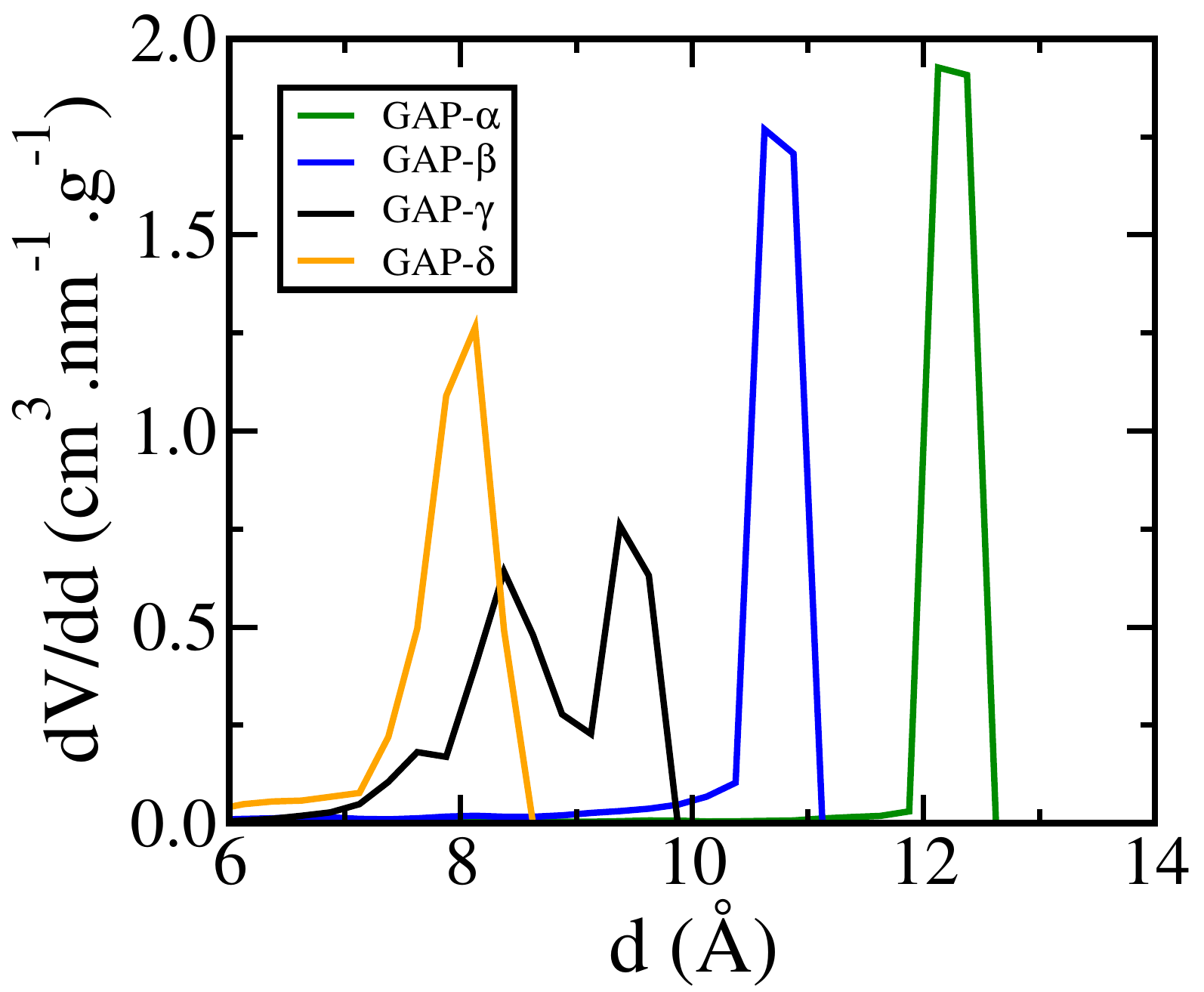}
\caption{Pore size distributions for the four carbons discussed in the main text, namely GAP-$\alpha$, GAP-$\beta$, GAP-$\gamma$ and GAP-$\delta$, with average pore sizes d$_{\rm avg}$= 12.1, 10.5, 8.7 and 7.7 \r{A}, respectively.}
\label{PSDs}
\end{figure}

In this study, we consider 10 different PSDs corresponding to atomistic structures reported by Deringer~\emph{et~al.}~\cite{Deringer18}, which were obtained by quench molecular dynamics using a machine learning based force field. These structures, with pore sizes ranging from 7~\r{A} to 13~\r{A}, are named GAP for the Gaussian Approximated Potential approach adopted for the force field. Figure \ref{PSDs} shows the PSDs of the four GAP carbons discussed in the main text. Additional carbons with micropores in the same range of pore sizes were also considered to generate the electrode structures (see Figure S1 in Supporting Information).  

Once we have determined the structural features of the carbon lattice, we need to add a description of the adsorption profiles of the considered species. Thanks to its multi-scale nature, our lattice model can use data from molecular simulations, performed at the slit pore level, to represent the adsorption in our electrode-scale model. In this study, we use data from MDs of two tetrafluoroborate based electrolytes, the neat ionic liquid (IL) [BMI][BF$_4$] and the ACN-[BMI][BF$_4$] organic electrolyte (ACN stands for acetonitrile) and two hexafluorophosphate based electrolytes, [BMI][PF$_6$] and ACN-[BMI][PF$_6$], in contact with graphene-like electrodes~\cite{MerletMD2013}. The concentration of the acetonitrile-based electrolytes are equal to 1.5~M. From the calculated free energy profiles of these species, an integrated density of adsorbed ions is determined as: 
\begin{equation}
      \rho=\rho_0\int_0^L r \exp{(\frac{-E_f}{{\rm k_B}T})} dr         
\end{equation}
where $\rho_0$ is the bulk ion density, $L$ is the width of the pore, ${\rm k_B}$ is the Boltzmann constant and $T$ is the temperature. In the main text, we focus on the BF$_4$-based electrolytes (the equivalent figures for the [BMI][PF$_6$]-based electrolytes are given in Supporting Information).  

\begin{figure}[ht!]
\centering
\includegraphics[scale=0.2]{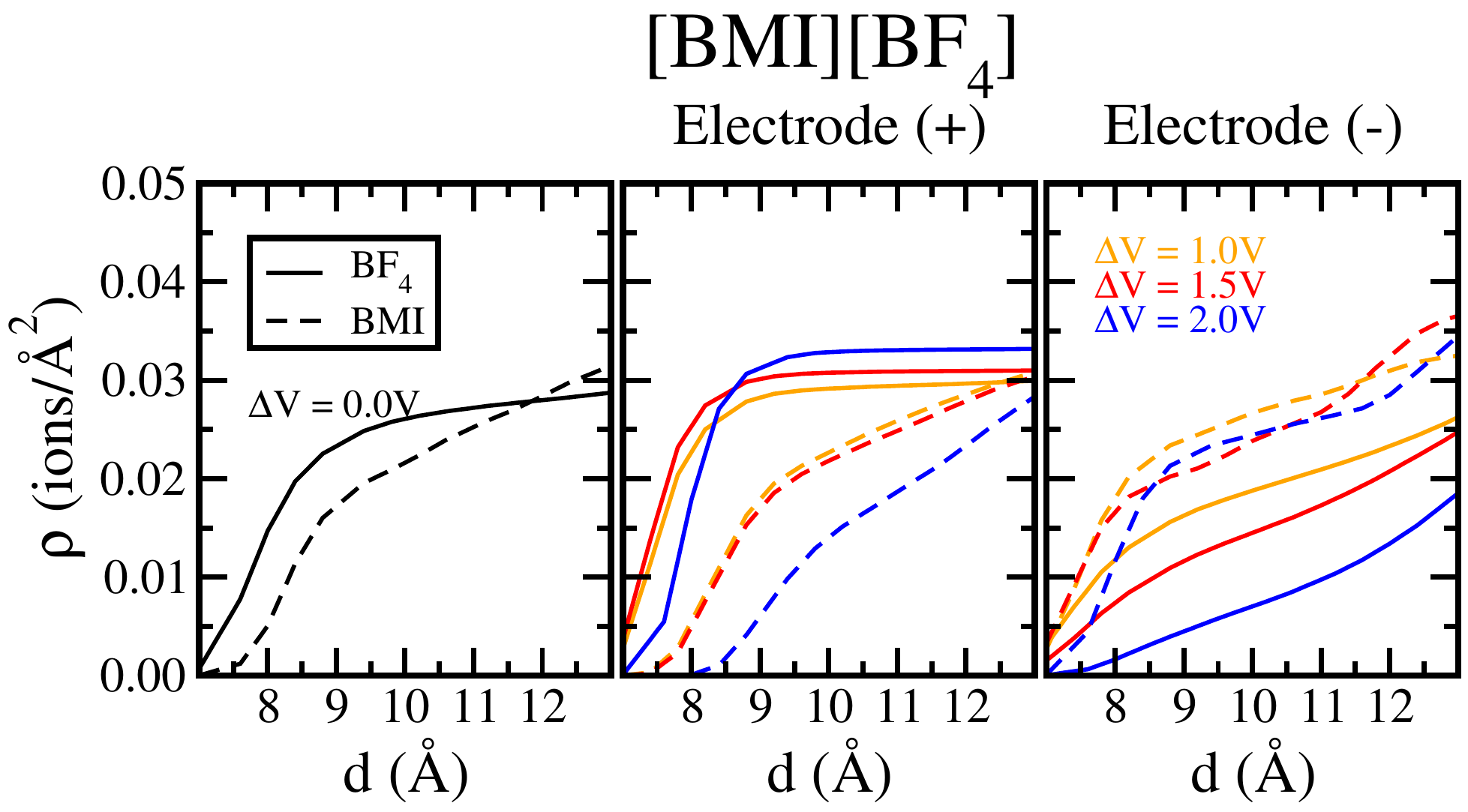}
\caption{Integrated density profiles of BF$_4^-$ anions (solid lines) and BMI$^+$ cations (dashed lines) in the [BMI][BF$_4$] neat IL as a function of pore size. The profiles are calculated for an applied potential difference of 0V, 1V, 1.5V and 2V.}
\label{intdens-BMIBF4}
\end{figure}

\begin{figure}[ht!]
\centering
\includegraphics[scale=0.2]{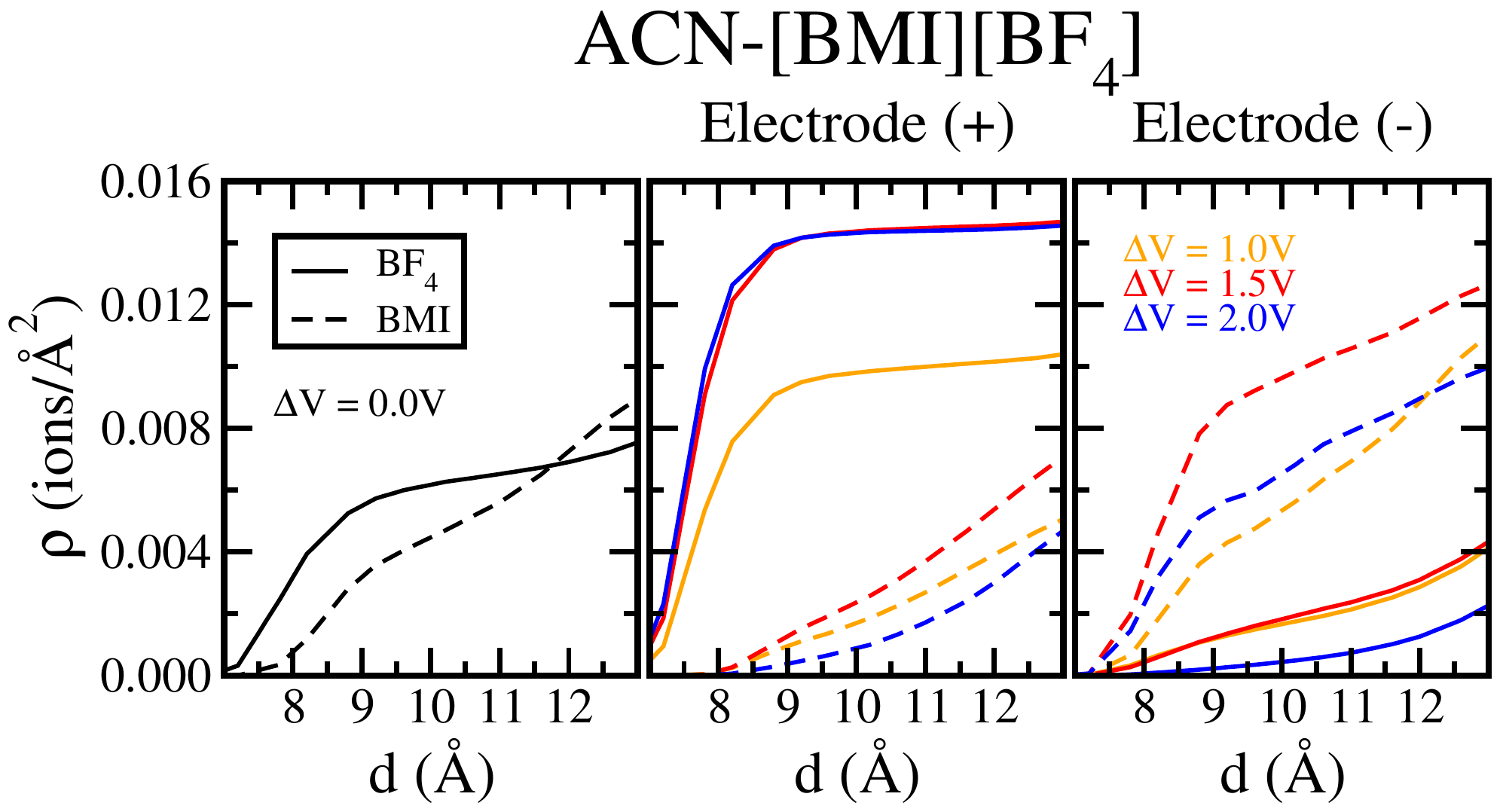}
\caption{Integrated density profiles of BF$_4^-$ anions (solid lines) and BMI$^+$ cations (dashed lines) in the ACN-[BMI][BF$_4$] organic electrolyte as a function of the pore size. The profiles are calculated for an applied potential difference of 0V, 1V, 1.5V and 2V.}
\label{intdens-ACNBMIBF4}
\end{figure}

Figures \ref{intdens-BMIBF4} and \ref{intdens-ACNBMIBF4} give the integrated densities as a function of the pore size for BF$_4^-$ anions and BMI$^+$ cations in the organic electrolyte and neat ionic liquid. The adsorbed densities are calculated for the case of a zero potential difference ($\Delta V$= 0V), as well as for applied potential differences of $\Delta V$= 1, 1.5 and 2V.    

From the integrated density profiles, we can see that, depending on their dimensions and on the potential applied, lattice sites are not identically populated. Therefore, each pore within the structure will have a different site energy $E_i$. This energy is equal to $E_i=-{\rm k_B} T \ln({\rho_i})$, with $\rho_i$  being the integrated density of site $i$, and conditions the lattice sites populations. To allow for ion diffusion between sites, we apply an acceptance rule to regulate inter-site transitions. The probability of a transition from site $i$ to site $j$ follows:

\begin{equation}
P(i,j) = \left\{
    \begin{array}{ll}
        \exp{(\frac{-(E_j - E_i)}{{\rm k_B} T})} & \mbox{if } E_j > E_i \\
         1 & \mbox{if } E_j \leq E_i
    \end{array}
\right.
\end{equation}

This condition is defined to favor jumps from sites with higher energies to sites with lower energies. A transition from site $i$ with higher energy to site $j$ with lower energy will always occur, while the probability of the opposite jump will decrease as the difference $E_i-E_j$ increases.

To run a lattice model simulation, a number of iterations $n$ and a timestep $\Delta t$ are defined. For each iteration, particles are moved using the moment-propagation scheme~\cite{Frenkel1987,Levesque2013,Rotenberg2008}, which is a recursive method allowing the examination of all possible trajectories at once and at a much lower computational cost than non-recursive methods. In the current state of the mesoscopic model, one calculation for a given system with 8,000 pores, i.e. corresponding to a 20$\times$20$\times$20 cubic lattice, takes around 2 to 3 hours on a single core. 

Inter-pore diffusion coefficients follow equation~\cite{Merlet15}:
\begin{equation}
    D_{ij}=\alpha_{ij}\frac{a^2}{2d\Delta t}
\end{equation}
where $a$ is the lattice spacing and $d$ is the dimensionality of the system. $\alpha_{ij}$ corresponds to a reduction factor for the inter-pore diffusion between two neighbouring lattice sites $i$ and $j$ compared to the bulk diffusion. This factor is defined as: $\alpha_{ij}=exp(\frac{-E_a(ij)}{{\rm k_B} T})$, where $E_a(ij)$ is the energy barrier governing jumps between lattice sites $i$ and $j$. The parametrisation of these energy barriers will be discussed in a later section.

\section{Results and discussion}

\subsection{In-pore ion populations}

\begin{figure}[ht!]
\centering
\includegraphics[scale=0.2]{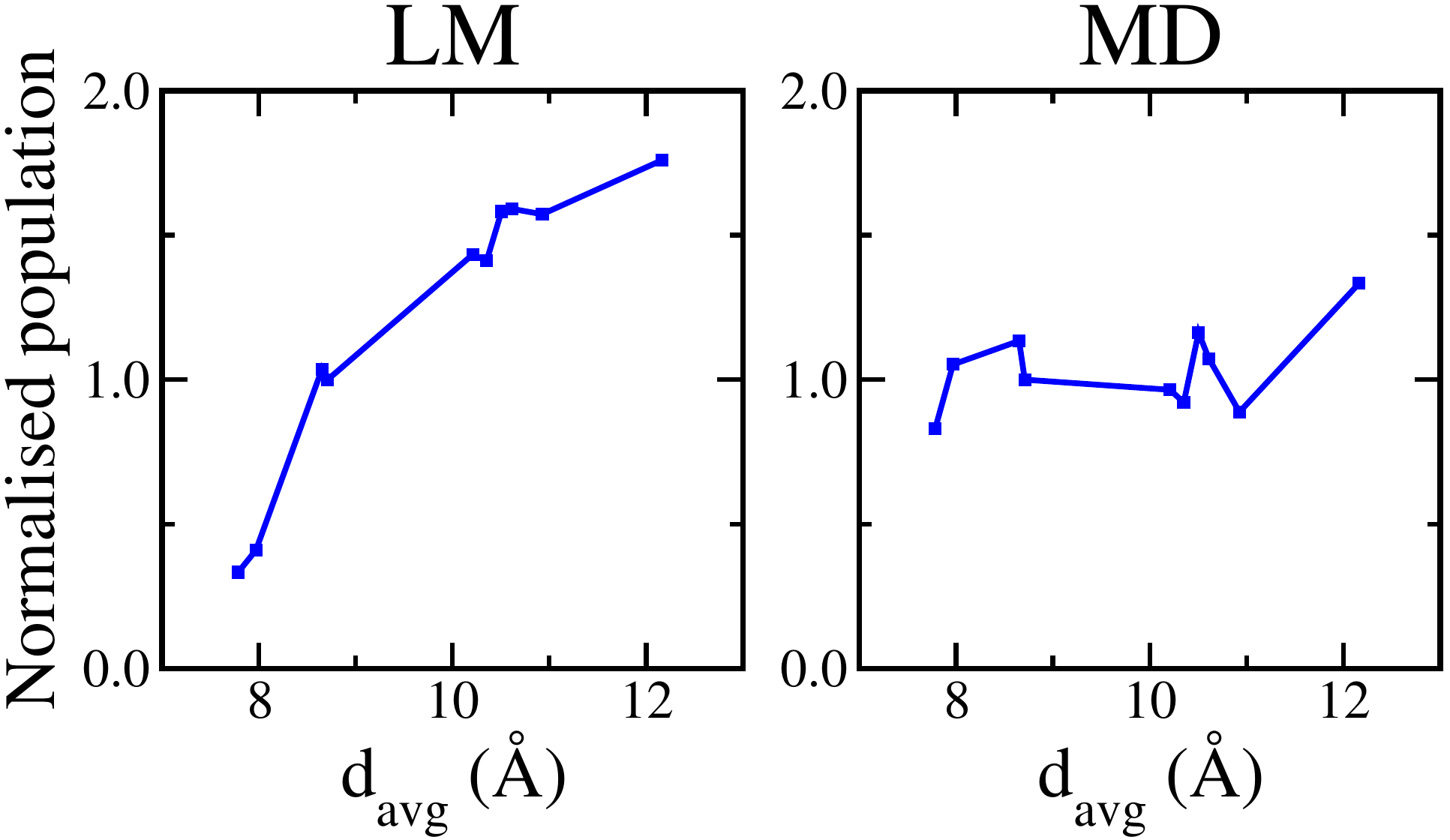}
\caption{Total in-pore population in GAP-based systems in contact with pure [BMI][PF$_6$]: comparison between the lattice model (LM) and Molecular Dynamics (MD) simulations. The calculations are carried out for an applied potential difference of 0V. Populations are normalised by the GAP carbon with d$_{\rm avg}$= 8.7 \r{A} (GAP-01 in the naming system of Deringer~\emph{et~al.}~\cite{Deringer18}.}
\label{LMMD}
\end{figure}

To examine the validity of our model we first compare the total in-pore ion populations, i.e. the sum of the number of adsorbed cations and the number of adsorbed anions  $N_{in-pore}=N^{BF_4}+N^{BMI}$, calculated within the lattice model to the ones obtained in the framework of MD simulations. The details of the simulations conducted are given in the Supporting Information. The comparison is done for all GAP electrodes in contact with neat [BMI][PF$_6$] at a 0V potential difference. The total in-pore populations are plotted as a function of the average pore size of the GAP carbons (see Figure \ref{LMMD}). To ease the comparison between the lattice model and the molecular simulations results, the populations are normalised by the population of an arbitrarily chosen GAP carbon, here GAP-01 corresponding to d$_{\rm avg}$= 8.7~\r{A}. The population profile obtained from lattice model simulations presents similar features as the one of MD simulations but overestimates the quantities for the largest pore sizes. An interesting result is that the lattice model is able to reproduce the peaks of populations observed at around 8.7~\r{A} and 10.6~\r{A}. This shows that our model, which operates at the electrode scale, can grasp some of the microscopic characteristics of the considered porous carbon-based supercapacitors. The overestimation could be due to the fact that our model underestimates the rugosity of the carbon structure, especially for pore interconnections. In the future, this could be improved by providing more accurate free energy profiles   

\begin{figure}[ht!]
\centering
\includegraphics[scale=0.5]{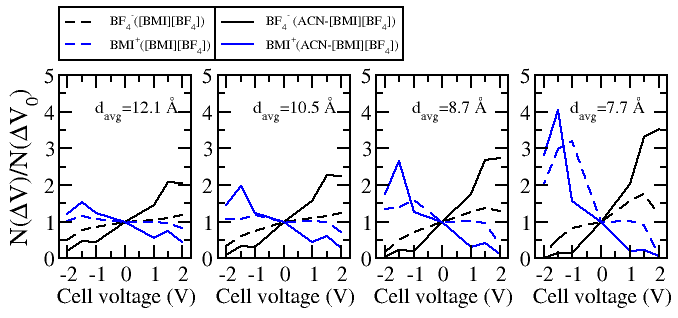}
\caption{Total in-pore ion population as a function of cell voltage for the considered electrode structures. Black and blue dashed lines show the populations of anions and cations, respectively, in the neat IL. Black and blue solid lines show the populations of anions and cations, respectively, in the organic electrolyte. Anions and cations populations are normalised by their equivalents at $\Delta V_0$= 0V. The voltages indicated correspond to the full cell voltage, e.g. 2V (respectively -2V) designates the positive electrode (respectively the negative electrode) for a 2V potential difference.}
\label{Norm_N}
\end{figure}

 We now focus on the evolution of the total in-pore population when charging the electrode. Figure \ref{Norm_N} compares the adsorption profiles of the neat ionic liquid and the organic electrolyte in four carbon structures. The charge storage mechanism at the negative and positive electrodes involves adsorption of counter-ions and desorption of co-ions, i.e. ionic exchange, for both [BMI][BF$_4$] and ACN-[BMI][BF$_4$], and that for all the considered carbon types. The populations in the charged electrodes are normalised by the values at $\Delta V_0$= 0V to examine the rate of adsorption/desorption mechanisms at non-zero applied potential differences. This rate shows a clear dependency on the pore size distributions, as well as on the absence/presence of solvent in the electrolyte. In general, when comparing carbons with different pore size distributions, we notice that higher ionic exchange occurs in carbons with smaller pores. 
 
To characterise more precisely the adsorption/desorption ratio, we quantify the ionic exchange rate as follows:  

\begin{equation}
    \alpha_{exchange}=\Bigl \lvert\frac{N^{BF_4}(\Delta V)}{N^{BF_4}(\Delta V_0)}-\frac{N^{BMI}(\Delta V)}{N^{BMI}(\Delta V_0)}\Bigr \rvert
\end{equation}

\begin{figure}[ht!]
\centering
\includegraphics[scale=0.5]{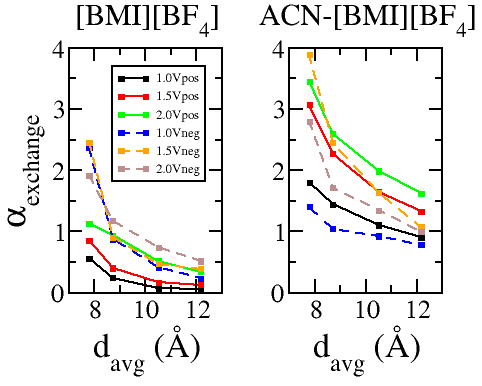}
\caption{Ionic exchange $\alpha_{exchange}$ as a function of the average pore size of the porous carbon structures. The exchange is calculated at the positive and negative electrodes for the considered applied potential differences.}
\label{exc}
\end{figure}

Figure \ref{exc} shows the ionic exchange rate $\alpha_{exchange}$ as a function of the average pore sizes of the carbons. For neat [BMI][BF$_4$], in addition to the decrease of ionic exchange for carbons with larger pore sizes, we notice that this mechanism is more pronounced at the negative electrode, compared to the positive one. This is observed for all the considered applied potentials (i.e. $\Delta V$= 1V, 1.5V and 2V). Since the anions and cations have different sizes, shapes and charge distributions, it is not surprising to observe an asymmetry between the positive and negative electrodes. The difference in mechanisms could be explained by a higher mobility of BMI$^+$ cations~\cite{These_Merlet}. At the negative electrode, applying a potential difference leads to the insertion of an important quantity of cations therefore forcing more BF$_4^-$ anions to exit the carbons micropores. This is clear at small pore sizes but not seen at larger pore sizes for which the relative differences between 0V and a non-zero potential difference are smaller. In contrast, at the positive electrode, the fast desorption of BMI$^+$ cations is accompanied by a slower adsorption of BF$_4^-$ anions in the considered voltage range, leading to a smaller ionic exchange. Again, this effect depends on the pore size as the large size of the cation probably leads to larger steric effects in the smallest pores.

Adding a solvent, here acetonitrile, to the IL greatly changes the dynamics of the in-pore ions adsorption. In this case, the solvent screens the electrostatic coupling between ions and increases the mobility of anions and cations within carbons pores. The solvent also stabilises the ions leading to an easier charge separation. This results in a notable increase of the ionic exchange, reaching in some cases three times the value of that of the neat IL. The solvent addition also introduces changes of the ionic orientations. This combined with the screening induced enhancement of ion mobility, which favors BF$_4^-$ anions in this case~\cite{These_Merlet}, influences the asymmetric character of the ionic exchange. In this case, the exchange rate tends to be higher at the positive electrode, as opposed to the previous case of the neat IL. These results are consistent with Electrochemical Quartz Crystal Microbalance experiments done on the [EMI][TFSI]-ACN electrolyte (1-ethyl-3-methylimidazolium  bis(trifluoromethanesulfonyl)imide)~\cite{Tsai14}. 

We note here that the calculations carried out for the 6~additional carbon structures and the [BMI][PF$_6$]-based electrolytes give similar results (see Figures S4 to S9 in Supporting Information). Moreover, the effect of the solvation leading to an increase of the ionic exchange was also observed in molecular simulations as described in Table S1 of the Supporting Information.

\subsection{Capacitive properties}

We pointed out that the charge storage mechanism involved simultaneous counter-ion adsorption and co-ion desorption in the carbon micropores. We now focus on the impact of the variations in ionic exchange on the capacitive properties of the electrodes considered. We calculate the absolute value of the total ionic charge stored in the electrode as: 
\begin{equation}
Q^{ionic}=\lvert N^{BF_4^-}-N^{BMI^+} \rvert e
\end{equation}
where $N^{BF_4^-}$ and $N^{BMI^+}$ are the in-pore populations of anions and cations, respectively, and $e$ is the elementary charge.

\begin{figure}[ht!]
\centering
\includegraphics[scale=0.48]{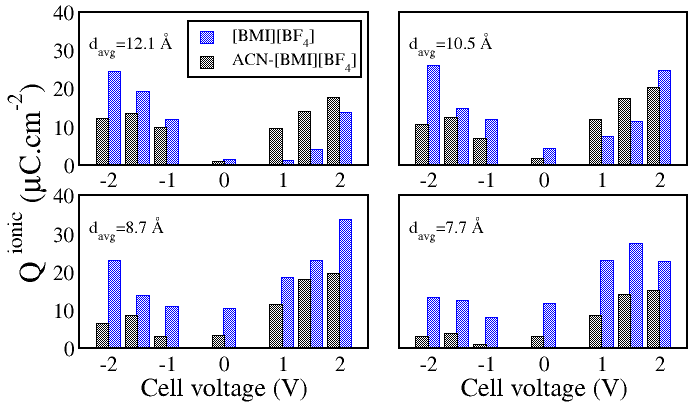}
\caption{Ionic charges stored at the positive and negative electrodes for the considered porous carbons. Blue and black bars give the charges obtained considering the neat IL and the organic electrolyte, respectively. The calculated ionic charges are normalised by the total electrode surface. The voltages indicated correspond to the full cell voltage, e.g. 2V (respectively -2V) designates the positive electrode (respectively the negative electrode) for a 2V potential difference.}
\label{Charge-V}
\end{figure}

Figure \ref{Charge-V} shows the calculated ionic charges for the considered carbon structures in contact with the neat IL and the organic electrolyte. The ionic charges normalised by the total surface area give values in the range of reported experimental data~\cite{Chmiola06,Largeot08}. We note that lattice simulations give non-zero stored charges at $\Delta V$= 0V. This results from the integrated ionic density profiles (see Figures~\ref{intdens-BMIBF4} and \ref{intdens-ACNBMIBF4}) extracted from MDs of electrolytes confined in larger slit pores than the pore sizes considered here. This is one aspect of the model which requires improvements, out of the scope of the current work, in order to give more realistic results.  

In general, systems containing the neat IL store considerably more charge than the ones with the organic electrolyte. This is surprising knowing that the ionic exchange occurs at a much higher rate using ACN-[BMI][BF$_4$]. However, the absolute numbers of adsorbed ions within the micropores are much larger in the neat IL than in the organic electrolyte (see Table S2 in Supporting information). This can explain, at least partially, the higher ionic charge in the neat IL. Moreover, for $\Delta V$= 0V, all carbon structures have a higher initial charge for the neat IL compared to the solvated electrolyte. Therefore, even though the rate of exchange is higher when charging the system containing the organic electrolyte, the higher initial charge for [BMI][BF$_4$] gives this electrolyte an advantage. 

The carbons with smaller micropores, i.e. GAP-$\gamma$ and GAP-$\delta$, further illustrate this trend. Indeed, for these structures the difference between [BMI][BF$_4$] and ACN-[BMI][BF$_4$] in the charge stored at $\Delta V$= 0V is much higher than for GAP-$\alpha$ and GAP-$\beta$. Upon charging, this difference is retained and a higher charge is stored for [BMI][BF$_4$] in contact with both the positive and negative electrode. For GAP-$\alpha$ and GAP-$\beta$, the smaller difference in charge storage at $\Delta V$=  0V is rapidly overtaken by the faster ionic exchange in the organic electrolyte, which appears to be dominant at the positive electrode. As a result, a higher charge is stored at the interface between the positive electrode and ACN-[BMI][BF$_4$].   

\begin{figure}[ht!]
\centering
\includegraphics[scale=0.44]{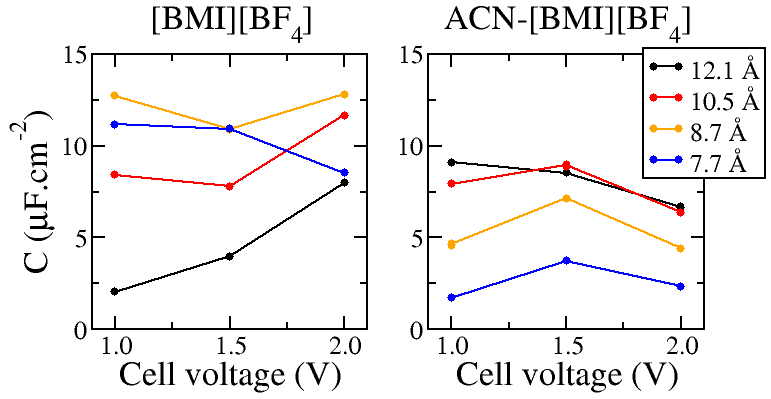}
\caption{Total capacitance as a function of the applied potential difference obtained for [BMI][BF$_4$] and ACN-[BMI][BF$_4$] in contact with GAP-$\alpha$, GAP-$\beta$, GAP-$\gamma$ and GAP-$\delta$. Capacitances are normalised by the total electrode surface.}
\label{Cap}
\end{figure}

From the obtained ionic charge we can calculate the capacitances of each electrode as: $C_{\pm}=Q^{ionic}_{\pm}/\Delta V_{\pm}$, with $C_+$ and $C_-$ representing the capacitances of the positive and negative electrode, respectively. The values for $\Delta V_{\pm}$, i.e. the potential drops at the electrode-electrolyte interface, are taken from MD simulations previously reported~\cite{MerletMD2013}. We note that the same trends are obtained if we simply take $\Delta V_+ = \Delta V_- = \Delta V/2$. Figure \ref{Cap} shows the total capacitance of the considered systems as a function of the applied potential difference. The total capacitance is calculated as:

\begin{equation}
    \frac{1}{C}=\frac{1}{C_+}+\frac{1}{C_-}
\end{equation}

We clearly see that the performance of the overall system is greatly influenced both by the carbon structure and the presence of solvent. When the electrode structure contains smaller pores, it shows better performances when in contact with the neat IL as the capacitance values in this case are clearly higher than the ones obtained with the organic electrolyte. As an example, for an applied potential difference of $\Delta V$= 1V, GAP-$\gamma$- and GAP-$\delta$-based systems (d$_{\rm avg}$= 8.7~\r{A} and d$_{\rm avg}$= 7.7~\r{A}, respectively) have a capacitance of 12.71 and 11.15~$\mu$F.cm$^{-2}$, respectively, when the system contains [BMI][BF$_4$], compared to 4.64 and 1.71~$\mu$F.cm$^{-2}$ for the ACN-[BMI][BF$_4$]-based systems. In contrast, the solvent addition improves the storage mechanism when the electrode contains larger pores. If we take the same case of $\Delta V$= 1V, we see that GAP-$\alpha$-based system (d$_{\rm avg}$= 12.1~\r{A}) present a higher capacitance for ACN-[BMI][BF$_4$] (9.08~$\mu$F.cm$^{-2}$) than for [BMI][BF$_4$] (2.04~$\mu$F.cm$^{-2}$). For GAP-$\beta$ (d$_{\rm avg}$= 10.5~\r{A}) the capacitive performance is improved when the system contains the organic electrolyte. Indeed, the difference in capacitance using ACN-[BMI][BF$_4$] and [BMI][BF$_4$] is considerably smaller (i.e. 7.92~$\mu$F.cm$^{-2}$ for ACN-[BMI][BF$_4$] and 8.42~$\mu$F.cm$^{-2}$ for [BMI][BF$_4$]), compared to the case of the electrode structure containing mainly small micropores.

It is quite surprising that the neat IL and the organic electrolyte give opposite variations with the pore size as this is not what is observed experimentally~\cite{Chmiola06,Largeot08}. One of the reason for this discrepancy could be the fact that in our lattice model, we represent only one electrode and as such, there is no interplay between the negative and positive electrodes which usually leads to one electrode being limiting in terms of charge storage. The difference in the stored ionic charges is an artefact of our model and is experimentally unrealistic in a two-electrode system. Strategies to overcome this limitation will be explored in a future work but comparisons with three-electrode systems might still be valid with our approach. Another reason for the unexpected variation with pore size could be the inaccuracy of the free energy profiles, this can be checked in future works by implementing a better description of these adsorption profiles. Finally, the experimental carbons are actually much more disordered than the carbons studied here, it is thus possible that such an effect would be observed in carbons with well defined pore sizes.

\subsection{Diffusion}

 As mentioned before, the diffusion coefficients of the ions and solvent mo\-le\-cu\-les in the lattice depend on the energy barriers between lattice sites, which are unknown. In previous works these energy barriers were assigned following a Gaussian distribution to calculate diffusion and NMR spectra in porous carbons~\cite{Merlet15}. However, this description lacks a dependency on the structural properties of the pores and of the adsorbed species. To overcome this, we now calculate energy barriers in order to reproduce experimental trends showing that diffusion coefficients decrease as the total in-pore population within the carbon structures increases~\cite{Forse17}.                  

\begin{figure}[ht!]
\centering
\includegraphics[scale=0.205]{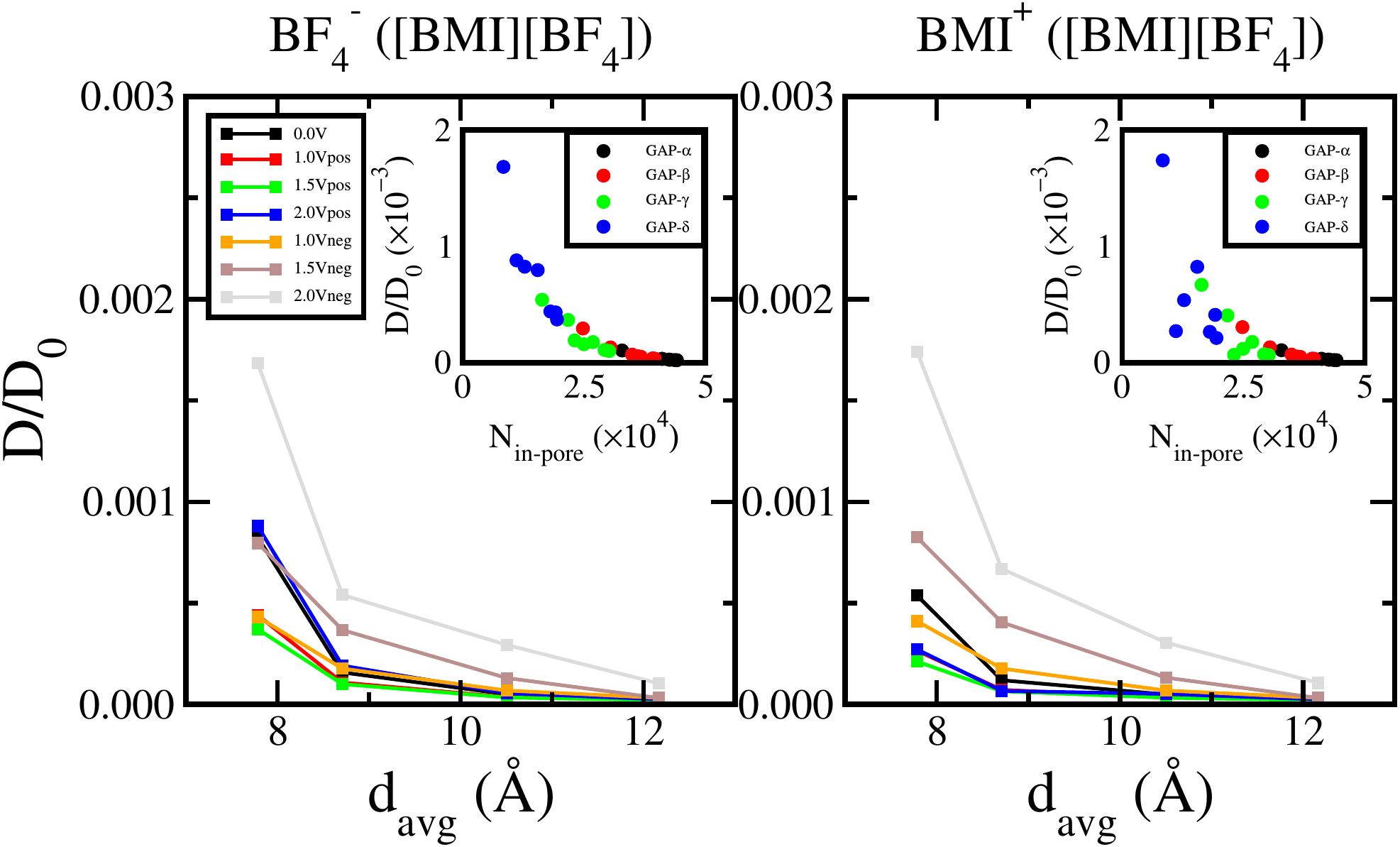}
\caption{Diffusion coefficients of adsorbed BF$_4^-$ anions and BMI$^+$ cations as a function of average pore size in systems containing the neat IL. Calculations are obtained for $\Delta V$= 0, 1, 1.5 and 2V. Insets: Diffusion coefficients as a function of the total in-pore population.}
\label{Diff1}
\end{figure}
\begin{figure}[ht!]
\centering
\includegraphics[scale=0.2]{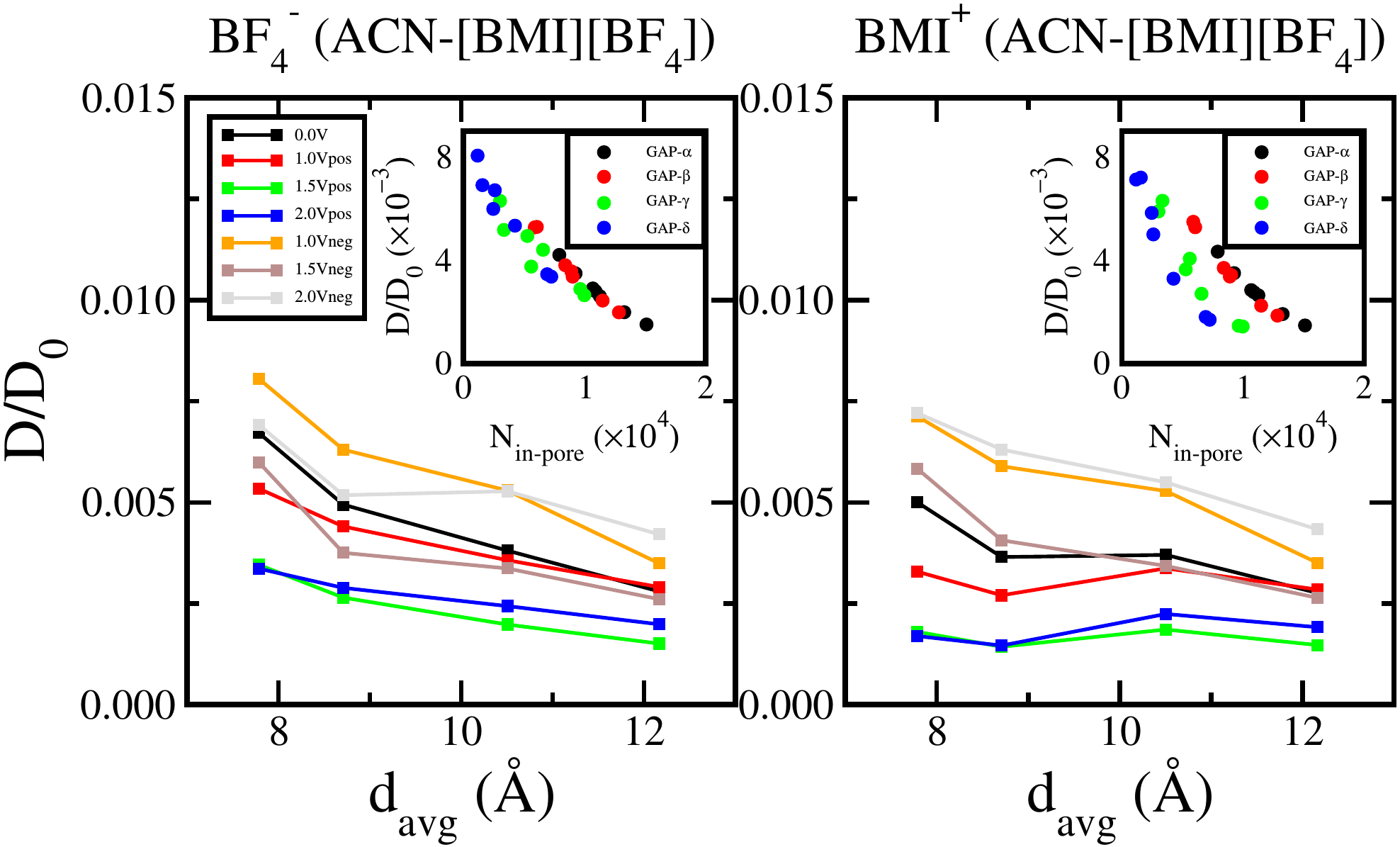}
\caption{Diffusion coefficients of adsorbed BF$_4^-$ anions and BMI$^+$ cations as a function of average pore size in systems containing the organic electrolyte. Results are shown for $\Delta V$= 0, 1, 1.5 and 2V. Insets: Diffusion coefficients as a function of the total in-pore population.}
\label{Diff2}
\end{figure}

The energy barriers considered to calculate the diffusion coefficients are fitted (using an exponential function) on the experimental data to reproduce the experimental decay of diffusion with the in-pore populations. Figures \ref{Diff1} and \ref{Diff2} show the resulting diffusion coefficients as a function of average pore sizes for various potentials and insets show their variation with the total in-pore populations. The results shown in the insets confirm that the energy barriers assigned allow us to reproduce the exponential decay with in-pore populations for the neat IL and for the organic electrolyte.

We now focus on the relationship between diffusion and structural properties of the considered carbon structures. Figures \ref{Diff1} and \ref{Diff2} give the diffusion coefficients (normalised by the bulk diffusion, $D_0=\frac{a^2}{2d\Delta t}$) as a function of the average pore sizes of the carbons for BF$_4^-$ and BMI$^+$ in [BMI][BF$_4$] and ACN-[BMI][BF$_4$] for all the considered applied potential differences.

In systems containing the organic electrolyte (figure \ref{Diff1}), we notice that diffusion tends to decrease in carbons with larger micropores for both anions and cations. Moreover, diffusion of a given ion type seems to occur at the same rate within both the positive and negative electrode. This dependency is in line with the ionic exchange profiles. When the pore size increases, the decay in diffusion is accompanied by a less dominant ionic exchange mechanism. The same dependencies are obtained when carbons are in contact with the neat IL (figure \ref{Diff2}), as the decrease of diffusion coefficients of BF$_4^-$ anions and BMI$^+$ cations is even more pronounced when considering carbons with larger micropores. 

Additionally, the solvation tends to amplify the inter-pore diffusion. Indeed, diffusion coefficients of anions and cations are almost one order of magnitude larger in systems with organic electrolyte compared to the ones with the neat IL. This is understandable knowing that solvation reduces the electrostatic coupling between ion pairs and the interaction of ions with pores walls. Moreover, this amplification of the diffusion induced by the solvation effect is also in agreement with the ionic exchange data, as we noted that the exchange mechanism occurs at a higher rate when considering the organic electrolyte compared to the neat IL. Finally, we remind that the dependencies discussed above were validated considering additional carbon structures and electrolyte compositions (see Figures S10 to S13 in the Supporting Information).

While the results obtained are consistent with our assignment of energy barriers, it is worth noting that we obtain larger diffusion coefficients for smaller pore sizes which is unexpected. Actually, the experimental results of Forse~\emph{et~al.}~\cite{Forse17} were obtained for a single carbon type and a single electrolyte. It is possible that the trend observed in not valid across a range of carbons. The variation of diffusion coefficients across carbon structures will be done in a future work using molecular simulations.

\section{Conclusions}

In this work, we report on the development and application of a mesoscopic lattice model used to study the relationship between structural and capacitive properties in porous carbon-based supercapacitors. This type of model was chosen for its high computational efficiency and we estimate that it is 10,000 times faster than molecular simulations. Several supercapacitor models were examined by considering various microporous carbon structures and electrolytes. In particular, the solvent effect was investigated by considering neat ILs, namely [BMI][BF$_4$] and [BMI][PF$_6$], and their acetonitrile-solvated equivalents. 

We showed that all systems exhibit ionic exchange upon charging the electrodes. This exchange appeared to be less important in carbon structures having large micropores, and more marked for organic electrolytes compared to neat ILs. The diffusion coefficients calculated follow this change of mechanism, with the decrease of diffusion coefficients in larger micropores and in neat ILs (compared to organic electrolytes). The lattice model also showed the effect of solvation on the charge storage in porous carbon-based supercapacitors. Even though ions exhibit higher exchange in organic electrolytes, the larger absolute numbers of adsorbed ions for neat ILs result in more charge storage for these systems. Moreover, the model showed that the capacitance tends to be higher in systems combining neat ILs or organic electrolytes with carbons containing mainly small or large micropores, respectively.  

In summary, our results show that the lattice model is able to retain some microscopic information and predict quantities of adsorbed ions, capacitances and diffusion coefficients in an efficient manner. However, for the lattice model to become suitable for a systematic study of the structural and dynamical properties of carbon-carbon supercapacitors, it is clear that further improvements are needed to better describe the microscopic information used as input. In particular the free energy profiles describing the ion adsorption and the energy barriers for diffusion. For instance, using data from molecular simulations performed on porous electrode structures, instead of the graphene-like systems considered here, could considerably improve the accuracy of the model. Other perspectives include the description of the free energy profiles through a classical density functional theory approach in order to remove the need for molecular simulations. 

\section*{Conflict of interests}

The authors declare no competing interest.

\section*{Data availability}

The data corresponding to the plots reported in this paper (pore size distributions, integrated densities, in-pore populations, ionic exchange values, capacitances, ionic charges and diffusion coefficients) are available in the Zenodo repository with identifier 10.5281/zenodo.3250139.

\section*{Acknowledgments}

This project has received funding from the European Research Council (ERC) under the European Union’s Horizon 2020 research and innovation programme (grant agreement no. 714581). This work was granted access to the HPC resources of CALMIP supercomputing center under the allocation P17037.

\end{document}